\documentclass[10pt,twoside]{amsart}
\usepackage[centertags]{amsmath}
\usepackage{amsfonts}
\usepackage[all]{xy}
\usepackage{epsfig}

\newcommand{\f}[2]{\frac{\displaystyle #1}{\displaystyle #2}}
\input{Qcircuit}
\def\sq{\sqrt}
\def\sq2{\sqrt{2}}
\def\sq12{\sq{12}}
\def\dsq2{\f{1}{\sqrt{2}}}
\def\be{\begin{equation}}
\def\ee{\end{equation}}
\def\lra{\longrightarrow}
\def\la{\langle}
\def\ra{\rangle}

\def\q={\quad = \quad}

\def\C{\mathbb{C}}

\newcommand{\vw}[2]{\left[ \begin{array}{c} #1 \\ #2 \end{array} \right]}

\catcode`\@=11\relax
\newdimen\p@renwd
\font\tenex=cmex10
\setbox0=\hbox{\tenex B}
\p@renwd=\wd0
\def\bbordermatrix#1{\begingroup \m@th
\setbox\z@\vbox{\def\\{\crcr\noalign{\kern2\p@\global\let\cr\endline}}%
    \ialign{$##$\hfil\kern2\p@\kern\p@renwd&\thinspace\hfil$##$\hfil
      &&\quad\hfil$##$\hfil\crcr
      \omit\strut\hfil\crcr\noalign{\kern-\baselineskip}%
      #1\crcr\omit\strut\cr}}%
  \setbox\tw@\vbox{\unvcopy\z@\global\setbox\@ne\lastbox}%
  \setbox\tw@\hbox{\unhbox\@ne\unskip\global\setbox\@ne\lastbox}%
  \setbox\tw@\hbox{$\kern\wd\@ne\kern-\p@renwd\left[\kern-\wd\@ne
    \global\setbox\@ne\vbox{\box\@ne\kern2\p@}%
    \vcenter{\kern-\ht\@ne\unvbox\z@\kern-\baselineskip}\,\right]$}%
  \null\;\vbox{\kern\ht\@ne\box\tw@}\endgroup}
\catcode`\@=12\relax
\begin{document}
\title{An Introduction to Quantum Computing}
\author{Noson S. Yanofsky}
\maketitle
\begin{abstract}
Quantum Computing is a new and exciting field at the intersection of mathematics, computer science and
physics. It concerns a utilization of quantum mechanics to improve the efficiency of computation.
Here we present a gentle introduction to some of the ideas in quantum computing. The paper begins by motivating the central ideas of
quantum mechanics and quantum computation with simple toy models. From there we move on to a formal presentation of the
small fraction of (finite dimensional) quantum mechanics that we will need for basic quantum computation.
Central notions of quantum architecture (qubits and quantum gates) are described. The paper ends with a
presentation of one of the simplest quantum algorithms: Deutsch's algorithm.
Our presentation demands neither
advanced mathematics nor advanced physics. \end{abstract}
\tableofcontents
\section[Intuition]{Intuition}

Quantum Computing is a fascinating new field at the intersection of
computer science, mathematics and physics. This field studies how to
harness some of the strange aspects of quantum physics for use in
computer science. Many of the texts to this field require
knowledge of a large corpus of advanced mathematics or physics.
We try to remedy this situation by presenting the basic ideas
of quantum computing understandable to anyone who has
had a course in pre-calculus or discrete structures. (A good course in linear
algebra would help, but, the reader is reminded of many definitions in the
footnotes.)

The reason why we are able to ignore the higher mathematics and physics is that
we do not aim to teach the reader all of quantum mechanics and all of quantum computing.
Rather, we lower our aim to simply present that part necessary to offer a taste of what
quantum computing is all about. What makes this possible is that we only need finite dimensional
quantum mechanics, i.e., the vector spaces that represent the states of the system will only
be of finite dimension. Such vector spaces consist of finite vectors with complex entries.
These vectors will change by being multiplied by operators or matrices. These matrices
will be finite and have complex entries. We do not do any more quantum computing than what is
needed for our final goal: Deutsch's algorithm. We stress that the reader does
not need more than the ability to do matrix multiplication in order to understand this paper.

To motivate our use of vectors to describe states and matrices as ways of
describing dynamics, we show that it is understandable if one looks at a basic toy models. Our models
deal with childrens' marbles moving along the edges of a graph.
Every computer scientist and logician
who has taken a class in discrete structures knows how to
represent a (weighted) graph as an adjacency matrix. We shall take
this basic idea and generalize it in several straightforward ways. While
doing this, we shall present many concepts that are at the very
core of quantum mechanics.

We begin with graphs that are without weights and progress to graphs that are weighted with real numbers, and finally
to graphs that are weighted with complex numbers.
With this in hand, we present a graph theoretic version of the double-slit experiment.
This is {\it the} most important experiment in quantum
mechanics. We conclude with a discussion of ways of combining systems to yield
larger systems.

Throughout this chapter, we shall present an idea in a toy model, then generalize it to an
abstract point and lastly discuss the connection with quantum mechanics before moving on to the
next idea.

This paper is based on a forthcoming text {\em Quantum Computing for Computer Scientists} coauthored with Mirco
Mannucci. The text was accepted for publication by Cambridge University Press and should see the light of day
in the beginning of 2008. In the text we take the reader through the same material and go much further.
The reader who appreciates this paper, will definitely gain from the text.
\\
\\
\noindent {\bf Acknowledgement.} I am grateful to Dr. Mirco Mannucci for many helpful discussions and cheery
editing sessions.

\subsection[Classical Deterministic Systems]{Classical Deterministic Systems}
We begin with a simple system described by a graph along with some children's marbles. Imagine the marbles
as being on the vertices of the graph. The state of a system is described by how many marbles are
on each vertex. For example, say that
there are six vertices in the graph and a total of 27 marbles.

We might place six marbles on vertex 0, two marbles on vertex 1 and the rest as described by this picture.
$$\xymatrix{
0 \bullet \fbox{\fbox{6}} && 1 \bullet  \fbox{\fbox{2}}&& \bullet 2\fbox{\fbox{1}}
 \\
3 \bullet \fbox{\fbox{5}}&& 4 \bullet \fbox{\fbox{3}} && \bullet 5 \fbox{\fbox{10}}}$$
We shall denote this state as $X=[6,2,1,5,3,10]^T$. The states of such a system will simply be
a collum vector of size 6.

We are not only interested in states of the system, but also in the way that the states
change --- or the ``dynamics'' of the system. This can be represented
by a graph with directed edges.
The dynamics might be described by the following directed graph:
$$\xymatrix{ 0 \bullet \ar[rrrrd]&& 1 \bullet \ar[rr] && \bullet 2 \ar[dll]
 \\
3 \bullet \ar@(r,u)[]  && 4 \bullet \ar[rr] && \bullet 5 \ar[u].}$$
The idea is that if there exists an arrow from vertex $i$ to vertex $j$, then in one
time click, all the marbles on vertex $i$
will move to vertex $j$. We place the following restriction on the types of graphs we shall
be concerned with: graphs with exactly one outgoing edge from each vertex. This will correspond
to the notion of a classical determinstic system. At each time click the marbles will have
exactly one place to go.

This graph is equivalent to the
matrix, $M$ (for ``marbles''):
$$ M \quad = \quad \  \bbordermatrix{
  & 0 & 1 & 2 & 3 & 4 & 5 \\
0 & 0 & 0 & 0 & 0 & 0 & 0 \\
1 & 0 & 0 & 0 & 0 & 0 & 0 \\
2 & 0 & 1 & 0 & 0 & 0 & 1 \\
3 & 0 & 0 & 0 & 1 & 0 & 0 \\
4 & 0 & 0 & 1 & 0 & 0 & 0 \\
5 & 1 & 0 & 0 & 0 & 1 & 0 \\
} $$
where $M[i,j]=1$ if and only if there is an arrow from vertex $j$ to vertex $i$.\footnote{Although most texts
might have $M[i,j]=1$ if and only if there is an arrow from vertex $i$ to vertex $j$,
we shall need it to be the other way for reasons which will become apparent later.
The difference is trivial.} Our restricted class of graphs are related to the restricted class of boolean
matrices that have exactly one 1 in each column.

Let us say we multiply $M$ by a state of the system $X=[6,2,1,5,3,10]^T$.  Then we have
$$ MX\quad=\quad \left[  \begin{array}{cccccc}
 0 & 0 & 0 & 0 & 0 & 0 \\
 0 & 0 & 0 & 0 & 0 & 0 \\
 0 & 1 & 0 & 0 & 0 & 1 \\
 0 & 0 & 0 & 1 & 0 & 0 \\
 0 & 0 & 1 & 0 & 0 & 0 \\
 1 & 0 & 0 & 0 & 1 & 0
 \end{array} \right]
\left[ \begin{array}{c} 6\\ 2\\ 1 \\5 \\ 3 \\10 \end{array} \right]
\quad = \quad
\left[ \begin{array}{c} 0 \\ 0 \\ 12\\5\\1\\9 \end{array} \right]\quad =\quad Y$$

What does this correspond to? If $X$ describes the state of the system at time $t$, then
$Y$ is the state at time $t+1$ , i.e., after one time click. We can see this clearly
by looking at the formula for matrix multiplication:
$$ Y[i]=(M\star X)[i] \quad = \quad \sum_{k=0}^5 M[i,k] X[k].$$
In plain English, this states that the number of marbles that will reach vertex $i$ after one time step
is the sum of all the marbles that are on vertices with edges connecting to vertex $i$.
Notice that the top two entries of $Y$ are zero. This corresponds to the fact that there are no arrows
going to vertex $0$ or $1$.

In general
if $X=[x_0, x_1, \ldots, x_{n-1}]^T$ is a column
vector corresponding to having $x_i$ marbles on vertex $i$, $M$ is a $n$ by $n$ Boolean matrix, and
if $MX=Y$ where $Y=[y_0, y_1, \ldots, y_{n-1}]^T$, then there are
$y_j$ marbles on vertex $j$ in one time
click. $M$ is thus a way of describing how the state of the marbles can
change from time $t$ to time $t+1$.

As we shall soon see, (finite dimensional) quantum mechanics works in the same way. States of a system are represented by
column vectors and the way in which the system changes in one time click is represented by matrices.
Multiplying a matrix with a column vector yields a new state of the system. Quantum mechanics explores
the way states of similar systems evolve over time.

\vspace{.5in}

Returning to our marbles, let's multiply $M$ by itself. $MM=M^2$.
However, since our entries are Boolean, we shall multiply the matrices
as Boolean, i.e., $1+1=1 \vee 1 =1$.
$$  \left[  \begin{array}{cccccc}
 0 & 0 & 0 & 0 & 0 & 0 \\
 0 & 0 & 0 & 0 & 0 & 0 \\
 0 & 1 & 0 & 0 & 0 & 1 \\
 0 & 0 & 0 & 1 & 0 & 0 \\
 0 & 0 & 1 & 0 & 0 & 0 \\
 1 & 0 & 0 & 0 & 1 & 0
 \end{array} \right]
 \left[  \begin{array}{cccccc}
 0 & 0 & 0 & 0 & 0 & 0 \\
 0 & 0 & 0 & 0 & 0 & 0 \\
 0 & 1 & 0 & 0 & 0 & 1 \\
 0 & 0 & 0 & 1 & 0 & 0 \\
 0 & 0 & 1 & 0 & 0 & 0 \\
 1 & 0 & 0 & 0 & 1 & 0
 \end{array} \right]
\q=
\left[  \begin{array}{cccccc}
 0 & 0 & 0 & 0 & 0 & 0 \\
 0 & 0 & 0 & 0 & 0 & 0 \\
 1 & 0 & 0 & 0 & 1 & 0 \\
 0 & 0 & 0 & 1 & 0 & 0 \\
 0 & 1 & 0 & 0 & 0 & 1 \\
 0 & 0 & 1 & 0 & 0 & 0
 \end{array} \right]
$$
Looking at the formula for Boolean matrix multiplication
$$M^2[i,j] \quad = \quad \bigvee_{k=0}^{n-1} M[i,k] \wedge M[k,j]$$
we can see that this formula really shows us how to go from vertex $j$ to vertex $i$ in
{\it two} time clicks.

And so we have that
$$ \begin{array}{rcl}
M^2[i,j]& = & \mbox{ 1 if and only if there is a path of length 2 from vertex }j \mbox{ to vertex } i. \\
\end{array} $$
For an arbitrary $k$ we have
$$ \begin{array}{rcl}
M^k[i,j]& = & \mbox{ 1 if and only if there is a path of length } k \mbox{ from vertex }j \mbox{ to vertex } i.
\end{array} $$

In general, multiplying an $n$ by $n$ matrix by itself several times will correspond to
the whether there is a path after several time clicks.
Consider $X=[x_0, x_1, \ldots , x_{n-1}]^T$ to be the state where
one has $x_0$ marbles on vertex 0, $x_1$ marbles on vertex 1,
$\ldots,$  $x_{n-1}$ marbles on vertex $n-1$. Then, after $k$ steps,
the state of the marbles is $Y$ where $Y=[y_0, y_1, \ldots,
y_{n-1}]^T=M^kX$. In other words, $y_j$ is the number of marbles on vertex $j$ after $k$ steps.

In quantum mechanics, if there are two or more matrices or operators that manipulate states, then
the action of one followed by another is described by their matrix product. We shall take
different states of systems and multiply the states by various matrices (of the appropriate type) to
obtain other states. These other states will again be multiplied by other matrices until we attain the
desired state.

\subsection[Classical Probabilistic Systems]{Classical Probabilistic Systems}
In quantum mechanics, neither the state of a system nor the dynamics of a system are deterministic.
There is an indeterminacy
in our knowledge of a state. Furthermore, the states change with probabilistic laws as opposed to deterministic laws.  That means that
states do not change in set ways. Rather, the laws
are given by stating that states will
change from one state to another state with a certain likelihood.

In order to capture these probabilistic scenarios, let us
generalize what we did in the last subsection. Instead of dealing with
a bunch of marbles moving around, we shall deal with a single marble.
The state of the system will tell us the probabilities of the single marble being on each vertex.
For a three-vertex graph, a typical state might look
like this $X=[\f{1}{5}, \f{3}{10},\f{1}{2}]^T$. This will correspond to the fact that there is
a one-fifth\footnote{Although the theory works with
any $r\in [0,1]$, we shall deal only with fractions.} chance that the marble is on vertex $0$;
a three-tenth chance that the marble is
on vertex $1$; and a half chance that the marble is on vertex $2$. Since the marble must be somewhere
on the graph, the sum of the probabilities is $1$.

We must generalize the dynamics as well.
Rather than exactly one arrow leaving each vertex, we will have
several arrows leaving each vertex with non-negative
real numbers between 0 and 1 as weights.
These weights will describe the probability of the single
marble going from one vertex to another in one time click. We shall restrict our
attention to weighted graphs that satisfy the following two
conditions: a) the sum of all the weights leaving a vertex is 1
and b) the sum of all the weights entering a vertex is 1.
This will correspond to the fact that a marble must go someplace
(there might be loops) and a marble must come from someplace. An
example of such a graph is

$$\xymatrix{ 0 \bullet \ar@/^/[rrrr]^{\f{1}{3}}\ar@/^/[rrdd]^{\f{2}{3}}
&&&& \bullet 1\ar@/^/[ddll]^{\f{1}{3}}\ar@/^/[llll]^{\f{1}{6}}
\ar@(r,u)[]_{\f{1}{2}}\\ \\
&& \bullet 2.\ar@/^/[uull]^{\f{5}{6}} \ar@/^/[uurr]^{\f{1}{6}}}$$
The matrix for this graph is
$$ M \quad = \quad \left[  \begin{array}{ccc}
0 & \f{1}{6} & \f{5}{6} \\
\f{1}{3} & \f{1}{2} & \f{1}{6} \\
\f{2}{3} & \f{1}{3} & 0 \end{array} \right] $$
The adjacency matrices for our graphs will have non-negative real entries where
the sums of the rows and the sums of the columns are all 1. Such matrices are
called ``doubly stochastic matrices.''

Let us see how the states interact with the dynamics. Suppose we have a state $X=\left[\f{1}{6}, \f{1}{6}, \f{2}{3}\right]^T$.
We will calculate how a state changes: $MX=Y$
$$ \left[  \begin{array}{ccc}
0 & \f{1}{6} & \f{5}{6} \\
\f{1}{3} & \f{1}{2} & \f{1}{6} \\
\f{2}{3} & \f{1}{3} & 0 \end{array} \right]
\left[ \begin{array}{c} \f{1}{6} \\ \f{1}{6} \\ \f{2}{3} \end{array} \right]
\quad = \quad
\left[ \begin{array}{c} \f{21}{36} \\ \f{9}{36} \\ \f{6}{36} \end{array} \right]$$
 Notice that the sum of the entries of $Y$ is 1.
If we have $X$ expressing the probability of the position of a marble, and $M$ expressing
the probability of the way the marble moves around,
then $MX=Y=\left[ \f{21}{36}, \f{9}{36}, \f{6}{36}\right]^T$
is to be interpreted as expressing the probability
of the marble's location after moving.
In other words, if $X$ is the probability of the marble at time $t$, then
$MX$ is the probability of the marble at time $t+1$.

If $M$ is an $n$ by $n$ doubly stochastic matrix and $X$ is an
$n$ by $1$ column vector whose entries sum to 1, then $M^kX=Y$
can be interpreted as expressing the probability of the position
of a marble after $k$ time clicks. That is, if $X=[x_0,x_1,
\ldots , x_{n-1}]^T$ means that there is an $x_i$ chance that a
marble is on vertex $i$, then $M^kX=Y=[y_0, y_1, \ldots,
y_{n-1}]^T$ means that after $k$ time clicks, there is a $y_j$
chance that the marble is on vertex $j$.

We are not constrained to multiply $M$ by itself. We may also multiply $M$ by another
doubly stochastic matrix. Let $M$ and $N$ be two $n$ by $n$ doubly stochastic
matrices. $M\star N$ will then describe a probability transition of going from
time $t$ to $t+1$ to $t+2$.

In quantum computers, quantum systems are generally in a probabilistic state. Manipulating the
system will correspond to multiplying the state by matrices. Each time click will correspond to
one matrix multiplication. At the end of the computation, the resulting vector will describe the
state of the system.

Before moving on to the next section, we shall examine an interesting
example. This shall be known as the ``probabilistic double slit experiment.''
Consider the following picture of a shooting gun.

\begin{figure}[htb]
\centering
\includegraphics[width=\textwidth, bb=0 0 800 600]{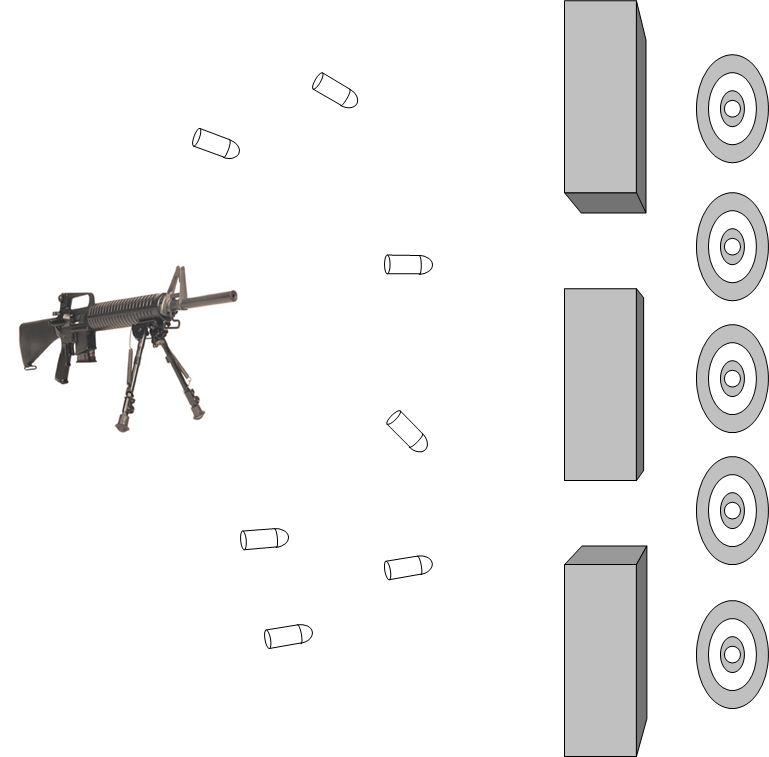}
\caption{Double slit experiment with bullets.}
\end{figure}

There are two slits in the wall. The shooter is a good enough shot to always get
the bullets through one of the two slits. There is a 50-50 chance of which slit the bullet will go through.
Once a bullet is through a slit, there are three targets to the right of each slit
that the bullet can hit with equal probability. The middle target can get
hit in one of two ways: from the top slit going down, or from the bottom slit going up. It is assumed that
it takes the bullet one time click to go from the gun to the wall and one time click to go from the wall to the
targets.
The picture correspond to the following weighted graph.
 $$\xymatrix{ &&&&\bullet 3
\ar@(r,u)[]_1
\\
\\
&& 1 \bullet \ar[rruu]^{\f{1}{3}} \ar[rr]^{\f{1}{3}} \ar[rrdd]^{\f{1}{3}}&& \bullet 4 \ar@(r,u)[]_1
\\
\\
0 \bullet \ar[rruu]^{\f{1}{2}}  \ar[rrdd]_{\f{1}{2}} &&&&  \bullet 5 \ar@(r,u)[]_1
\\
\\
&& 2 \bullet \ar[rruu]^{\f{1}{3}} \ar[rr]^{\f{1}{3}} \ar[rrdd]^{\f{1}{3}}&& \bullet 6 \ar@(r,u)[]_1
\\
\\
&&&&\bullet 7 \ar@(r,u)[]_1}$$
Notice that the vertex marked 5 can receive bullets from either
of the two slits.

Corresponding to this graph is the matrix $B$ (for ``bullets'')
$$B= \left[ \begin{array}{cccccccc}
0&0&0&0&0&0&0&0\\
\f{1}{2}&0&0&0&0&0&0&0\\
\f{1}{2}&0&0&0&0&0&0&0\\
0&\f{1}{3}&0&1&0&0&0&0\\
0&\f{1}{3}&0&0&1&0&0&0\\
0&\f{1}{3}&\f{1}{3}&0&0&1&0&0\\
0&0&\f{1}{3}&0&0&0&1&0\\
0&0&\f{1}{3}&0&0&0&0&1
\end{array} \right]$$

In words, $B$ describes the way a bullet will move after one time click.\footnote{
The matrix $B$ is not a doubly
stochastic matrix. The sum of the weights entering vertex 0 is
not 1. The sum of weights leaving vertices 3, 4, 5, 6, and 7 are more than 1. This
fact should not bother you. We are interested in demonstrating the way
probabilities behave with respect to these matrices.}

Let us calculate the probabilities for the bullet's position after two time clicks.
$$B\star B \q= B^2\q= \left[ \begin{array}{cccccccc}
0&0&0&0&0&0&0&0\\
0&0&0&0&0&0&0&0\\
0&0&0&0&0&0&0&0\\
\f{1}{6}&\f{1}{3}&0&1&0&0&0&0\\
\f{1}{6}&\f{1}{3}&0&0&1&0&0&0\\
\f{1}{3}&\f{1}{3}&\f{1}{3}&0&0&1&0&0\\
\f{1}{6}&0&\f{1}{3}&0&0&0&1&0\\
\f{1}{6}&0&\f{1}{3}&0&0&0&0&1
\end{array}\right]$$
So $B^2$ indicates the probabilities of the bullet's position after two time clicks.

If we are sure that we start with the bullet in position 0, i.e., $X=[1,0,0,0,0,0,0,0]^T$,
then, after two time clicks, the state of the bullets will be
$$ B^2X=[ 0, 0, 0, \f{1}{6}, \f{1}{6}, \f{1}{3}, \f{1}{6}, \f{1}{6}]^T$$
The key idea is to notice that $B^2[5,0]=\f{1}{6}+\f{1}{6}=\f{1}{3}$
because the bullets start from position 0, then there are two
possible ways of the bullet getting to position 5. The
possibilities sum to $\f{1}{3}$. This is what we would expect. We
shall revisit this example in the next subsection where strange things
start happening!

\subsection[Quantum Systems]{Quantum Systems}
We are now ready to leave the world of classical probabilities and enter the world of the
quantum. One of the central facts about quantum mechanics is that complex numbers play a major role in the calculations.
Probabilities of states and transitions are not given as a real numbers $p$ between 0 and 1.
Rather, they are given is a complex numbers $c$ such that $|c|^2$ is a real number\footnote{We remind the reader that if $c=a+bi$ is a
complex number, then its modulus is $|c|=\sqrt{a^2+b^2}$ and $|c|^2=a^2+b^2$.} between 0 and 1.

What is the difference how the probabilities are given? What does it matter
if a probability is given as a real number between 0 and 1, or as a complex number whose
modulus squared is a real number between 0 and 1? The difference is --- and this is the core of quantum
theory --- that real number
probabilities can be added to obtain larger real numbers. In contrast, complex numbers can cancel
each other and lower their probability. In detail, if $p_1$ and $p_2$ are two real numbers between
0 and 1, then $(p_1 + p_2)\geq p_1$ and $(p_1 + p_2)\geq p_2$. Now let's look at the complex case.
Let $c_1$ and $c_2$ be two complex numbers with their squares of modulus $|c_1|^2$ and
$|c_2|^2$. $|c_1 + c_2|^2$ need not be bigger then $|c_1|^2$ and it also does not need to be bigger than $|c_2|^2$.

For example\footnote{The important point here is that the modulus squared is positive.
For simplicity of calculations, we have chosen easy complex numbers.}
, if $c_1=5+3i$ and $c_2= -3 -2i$, then $|c_1|^2=34$ and $|c_2|^2=13$ but
$|c_1+c_2|^2 = | 2+i|^2 = 5$. $5$ is less than $34$ and $5$ is less than $13$.

This possibility of canceling out complex numbers corresponds to something called ``interference'' in quantum mechanics.
One complex number might interfere with another.
It is one of the most important ideas in quantum theory.

Let us generalize
our states and graphs from the previous subsection. Rather than insisting
that the sum of the entries in the column vector is 1, we insist that the sum of the modulus squared of the entries is 1.
This makes sense since we are considering the probability as the modulus squared.

For dynamics, rather than talking about graphs with real number weights, we shall
talk about graphs with complex number weights. Instead of insisting that the adjacency
matrix of such a graph
be a doubly stochastic matrix, we ask instead that the adjacency matrix be unitary.\footnote{Let us
just remember: a matrix $U$ is unitary if $U\star U^\dag = I = U^\dag \star U$. The adjoint of $U$, denoted as
$U^\dag$, is defined as $U^\dag=(\overline{U})^T= \overline{(U^T)}$ or $U^\dag [j,k] = \overline{U[k,j]}$.
}

For example, consider the graph $$ \xymatrix{ 0\bullet
\ar@(l,u)[]^{\f{1}{\sqrt{2}}} \ar@/^/[rrrr]^{\f{-i}{\sqrt{2}}}
&&&&
\bullet 1 \ar@/^/[llll]^{\f{1}{\sqrt{2}}} \ar@(u,r)[]^{\f{i}{\sqrt{2}}}\\ \\
&& \bullet 2. \ar@(r,u)[]_{i} }$$

The corresponding unitary adjacency matrix is
$$U \q= \left[  \begin{array}{ccc}
\f{1}{\sqrt{2}}& \f{1}{\sqrt{2}} & 0 \\
\f{-i}{\sqrt{2}} & \f{i}{\sqrt{2}} & 0 \\
0 & 0 & i \end{array} \right].$$

Unitary matrices are related to doubly stochastic matrices as follows.
The modulus squared of the all the complex entries in $U$ forms a
doubly stochastic matrix. The $i,j$th element in $U$ is denoted $U[i,j]$, and
its modulus squared is denoted $|U[i,j]|^2$. By abuse of notation, we
shall denote the entire matrix of modulus squares as $|U[i,j]|^2$:
$$|U[i,j]|^2 \q= \left[ \begin{array}{ccc}
\f{1}{2}& \f{1}{2}& 0\\
\f{1}{2}& \f{1}{2} & 0\\
0& 0& 1 \end{array} \right]. $$ It is easy to see that this is a doubly stochastic matrix.

From this graph-theoretic point of view, it is easy to see what unitary means:
the conjugate transpose of the $U$ matrix is
$$U^\dag \q= \left[  \begin{array}{ccc}
\f{1}{\sqrt{2}}& \f{i}{\sqrt{2}} & 0 \\
\f{1}{\sqrt{2}} & \f{-i}{\sqrt{2}} & 0 \\
0 & 0 & -i \end{array} \right].$$
This matrix corresponds to the graph $$ \xymatrix{ 0\bullet
\ar@(l,u)[]^{\f{1}{\sqrt{2}}} \ar@/^/[rrrr]^{\f{1}{\sqrt{2}}} &&&&
\bullet 1 \ar@/^/[llll]^{\f{i}{\sqrt{2}}} \ar@(r,u)[]_{\f{-i}{\sqrt{2}}}\\
&& \bullet 2. \ar@(r,u)[]_{-i} }$$
If $U$ is the matrix that takes a state from time $t$ to time $t+1$, then $U^\dag$ is the
matrix that takes a state from time $t$ to time $t-1$.
If we multiply $U$ and $U^\dag$, we get the identity matrix $I_3$ which corresponds
to the graph
$$ \xymatrix{ 0\bullet \ar@(r,u)[]_1 &&&&
\bullet 1 \ar@(r,u)[]_1 \\
&& \bullet 2. \ar@(r,u)[]_{1} }$$
This means that if we perform some operation and then ``undo'' the operation, we will
find ourselves in the same state as we began with probability 1. It is important to note that unitary does not only mean
invertible. It means invertible in a very easy way, i.e., the inverse is the dagger of the matrix. This ``invertibility'' is
again an important issue in quantum mechanics. Most of the dynamics will be invertible (except measurements).

\vspace{.5in}

In order to see the interference phenomenon and the related ``superposition'' phenomenon, we will revisit the
double-slit experiment from the last subsection. Rather than looking at bullets,
which are relatively large objects and hence adhere to the laws of
classical physics, we shall look at microscopic objects such as
photons that follow the laws of quantum physics. Rather than having a gun, we shall have a laser shoot
photons through two slits as in figure 2.

\begin{figure}[htb]
\centering
\includegraphics[width=\textwidth, bb=0 0 800 600]{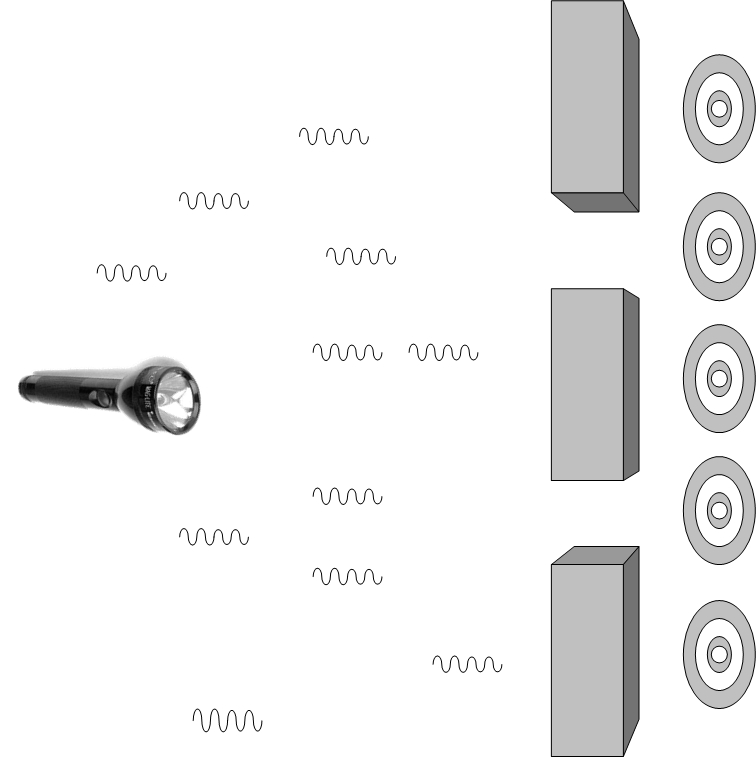}
\caption{Double slit experiment with photons.}
\end{figure}

Again we shall make the assumption that a photon will pass through
one of the two slits. Each slit has a 50\% chance of the photon
going through it. To the left of each slit there are three measuring
devices. It is assumed that it takes one time click to go from the
laser to the slits and one time click to go from the slits to the
targets. We are not interested in how large the slits are or how far
the measuring positions are from the slits. Physicists are very
adapt at calculating many different aspects of this experiment. We
are only interested in the set-up.

Consider the following weighted graph that describes the set-up of the experiment:
$$\xymatrix{
&&&&\bullet 3 \ar@(r,u)[]_1
\\
\\
&& 1 \bullet \ar[rruu]^{\f{-1+i}{\sqrt{6}}} \ar[rr]^{\f{-1-i}{\sqrt{6}}} \ar[rrdd]_{\f{1-i}{\sqrt{6}}}&&
\bullet 4 \ar@(r,u)[]_1
\\
\\
0 \bullet \ar[rruu]^{\dsq2}  \ar[rrdd]_{\dsq2} &&&&  \bullet 5\ar@(r,u)[]_1
\\
\\
&& 2 \bullet \ar[rruu]^{\f{-1+i}{\sqrt{6}}} \ar[rr]^{\f{-1-i}{\sqrt{6}}} \ar[rrdd]_{\f{1-i}{\sqrt{6}}}&&
\bullet 6 \ar@(r,u)[]_1
\\
\\
&&&&\bullet 7. \ar@(r,u)[]_1}$$
The modulus squared of $\f{1}{\sqrt{2}}$ is $\f{1}{2}$, which corresponds to the fact that
there is a 50-50 chance of the photon going through either slit. $\left| \f{\pm 1 \pm i }{\sqrt{6}}\right|^2
= \f{1}{3}$ which corresponds to the fact that whichever slit the photon goes through, there is a $\f{1}{3}$
of a chance of its hitting any of the three measuring positions to the right of that slit.
\footnote{The actual complex number weights are not our interest here.  If we wanted to calculate the actual numbers,
we would have to measure the width of the
slits, the distance between the slits, the distance from the slits to the measuring devices etc. However, our goal here
is to clearly demonstrate the interference phenomenon. And so we chose the above complex numbers simply
because the modulus squared are
exactly the same as the bullets case.}

The adjacency matrix for this graph is $P$ (for ``photons'')\footnote{This matrix is not a unitary matrix.
Looking carefully at row 0, one can immediately see that $P$ is not unitary.
In our graph, there is nothing entering vertex 0.
The reason why this matrix fails to be unitary is because we have not put in all the arrows in our graph. There are
many more possible ways the photon can travel in a real-life physical situation. In particular,
the photon might go from the right to
the left. The diagram and matrix would become too complicated if we put in all the transitions.
We are simply trying to demonstrate the interference phenomenon and we can
accomplish that even with a matrix that is not quite unitary.}
$$P \q= \left[ \begin{array}{cccccccc}
0&0&0&0&0&0&0&0\\
\dsq2&0&0&0&0&0&0&0\\
\dsq2&0&0&0&0&0&0&0\\
0&\f{-1+i}{\sqrt{6}}&0&1&0&0&0&0\\
0&\f{-1-i}{\sqrt{6}}&0&0&1&0&0&0\\
0&\f{1-i}{\sqrt{6}}&\f{-1+i}{\sqrt{6}}&0&0&1&0&0\\
0&0&\f{-1-i}{\sqrt{6}}&0&0&0&1&0\\
0&0&\f{1-i}{\sqrt{6}}&0&0&0&0&1
\end{array} \right].$$

The modulus squared of the $P$ matrix is exactly the same as the bullets matrix i.e.,
$|P[i,j]|^2 \q= B$. Let us see what happens if we calculate
the transitions matrix after {\em two} time clicks.
$$P^2 \q= \left[ \begin{array}{cccccccc}
0&0&0&0&0&0&0&0\\
0&0&0&0&0&0&0&0\\
0&0&0&0&0&0&0&0\\
\f{-1+i}{\sqrt{12}}&\f{-1+i}{\sqrt{6}}&0&1&0&0&0&0\\
\f{-1-i}{\sqrt{12}}&\f{-1-i}{\sqrt{6}}&0&0&1&0&0&0\\
0&\f{-1+i}{\sqrt{6}}&\f{1-i}{\sqrt{6}}&0&0&1&0&0\\
\f{-1-i}{\sqrt{12}}&0&\f{-1-i}{\sqrt{6}}&0&0&0&1&0\\
\f{-1+i}{\sqrt{12}}&0&\f{1-i}{\sqrt{6}}&0&0&0&0&1
\end{array} \right].$$

How do we  interpret this in terms of probability? Let us look at the modulus squared
of each of the entries.
$$|P^2[i,j]|^2 \q= \left[ \begin{array}{cccccccc}
0&0&0&0&0&0&0&0\\
0&0&0&0&0&0&0&0\\
0&0&0&0&0&0&0&0\\
\f{1}{6}&\f{1}{3}&0&1&0&0&0&0\\
\f{1}{6}&\f{1}{3}&0&0&1&0&0&0\\
{\Huge \bf 0}&\f{1}{3}&\f{1}{3}&0&0&1&0&0\\
\f{1}{6}&0&\f{1}{3}&0&0&0&1&0\\
\f{1}{6}&0&\f{1}{3}&0&0&0&0&1
\end{array}\right].$$

This matrix is almost exactly the same as $B^2$
but with one glaring difference. $B^2[5,0] = \f{1}{3}$ because of
the two ways of starting at position 0 and ending at position 5.
We added the nonnegative probabilities $\f{1}{6} + \f{1}{6} =
\f{1}{3}$. However with a photon that follows the laws of quantum
mechanics, the complex numbers are added as opposed to their
probabilities. $$\dsq2\left(\f{-1+i}{\sqrt{6}}\right) +
\dsq2\left(\f{1-i}{\sqrt{6}}\right) \q= \f{-1+i}{\sqrt{12}} +
\f{1-i}{\sqrt{12}} \q= \f{0}{\sqrt{12}} \q= 0.$$ Thus giving us
$|P^2[5,0]|^2=0$. In other words, although there are two ways of
a photon going from vertex 0 to vertex 5, there will be no photon
at vertex 5.

How is one to understand this phenomenon? Physicists have a simple
explanation for interference: waves. A familiar observation such as a pebble thrown into a pool
of water will easily convince us that waves interfere, sometimes reinforcing each other,
sometimes cancelling each other. In our experiment, photons are going through both
slits at one time and they are canceling each other out at the middle measuring device.
Thus, the double-slit experiment points to the wave-like nature of light.


The experiment can be done with only one photon shot out from vertex 0. This ensures that there will not be another
wave for it to cancel out with. And yet, when only one photon goes through a slit, there is still
an interference phenomenon. What is going on here?

The naive probabilistic interpretation of the
position of the photon following the bullet metaphor of the previous subsection is thus
not entirely adequate. Let the state of the system be
given by $X=[c_0,c_1, \ldots, c_{n-1}]^T \in \mathbb{C}^n$. It is
incorrect to say that the probability of the photon being in position
$k$ is $|c_k|^2$. Rather, a system in state $X$ means that the
particle is in {\em all} positions simultaneously. It is only after we measure the
photon that the chances of it being found in position
$k$ is $|c_k|^2$. The photon (or its associated wave) goes through the
top slit {\em and} the bottom slit simultaneously. And when the photon exits both
slits, it can cancel {\em itself} out. A photon is not in {\em a}
singe position, rather it is in {\em many} positions or a ``superposition''.

This might cause some justifiable disbelief. After
all, we do not see things in a superposition of states. Our everyday
experience tells us that things are in one position or (exclusive
or!) another position. How can this be? The reason why we see particles in
exactly one position is because we have performed a
measurement. When we measure something at the quantum level, the quantum object that we
have measured is no longer in a superposition of states, rather it
collapses to a single classical state. So we have to redefine what a state of a quantum system means:
A system in state $X$ means that
{\em after measuring} the photon it will be in position $k$ with
probability $|c_k|^2$.

What are we to make of these strange ideas? Are we really to believe them?
Richard Feynman in discussing the double-slit experiment \cite{Feynman} (Vol. III, Page 1-1) waxes lyrical:
\begin{quote} We choose to
examine a phenomenon which is impossible, {\it absolutely} impossible, to explain in any classical way, and which
has in it the heart of quantum mechanics. In reality, it contains the {\it only} mystery. We can not make the mystery
go away by ``explaining'' how it works. We will just tell you how it works.
\end{quote}

It is exactly this superposition of states that is the real power behind quantum computing. Classical computers
are in one state at every moment. Quantum computers can be put in a superposition of states. Imagine
putting a computer in many different classical states at one time and then processing with {\em all} the states. This is
the ultimate in parallel processing! Such a computer can only be conceived in the quantum world.

\subsection[Combining Systems]{Combining Systems}
Quantum mechanics can also deal with systems that have more than
one part. In this subsection we learn how to combine several systems into one.
We shall talk about combining classical probabilistic systems. However, whatever is stated about classical probabilistic systems
is also true for quantum systems.

Consider two different marbles. Imagine that a red marble follows
the probabilities of the graph
whose corresponding adjacency matrix is
$$ M \quad = \quad \left[  \begin{array}{ccc}
0 & \f{1}{6} & \f{5}{6} \\
\f{1}{3} & \f{1}{2} & \f{1}{6} \\
\f{2}{3} & \f{1}{3} & 0 \end{array} \right]. $$
Consider also a blue marble that follows the transitions given by the graph
$$G_N \q= \xymatrix{a\bullet \ar@(l,u)[]^{\f{1}{3}} \ar@/^/[rrrr]^{\f{2}{3}}&&&&
\bullet b \ar@(r,u)[]_{\f{1}{3}} \ar@/^/[llll]^{\f{2}{3}}}$$
i.e., the matrix $$N \q= \left[ \begin{array}{cc}
\f{1}{3} & \f{2}{3} \\
\f{2}{3} & \f{1}{3} \end{array} \right]. $$

What does a state for a {\em two} marble system look like?
Since the red marble can be on one of three  vertices and the blue marble can be on
one of two vertices, there are $3 \times 2=6$ possible states of the combined system.
This is the tensor product of a 3 by 1 vector with a 2 by 1 vector. A typical state might
look like:
$$X \q= \bbordermatrix{ & \\
0a& \f{1}{18} \\
0b &0 \\
1a& \f{2}{18} \\
1b& \f{1}{3} \\
2a & 0 \\
2b &\f{1}{2}
},$$
which would correspond to the fact that there is a
\begin{verse}
$\f{1}{18}$ chance of the red marble being on vertex 1 and the blue marble being on vertex $a$, \\
$0$ chance of the red marble being on vertex 1 and the blue marble being on vertex $b$, \\
$\vdots$\\
$\f{1}{2}$ chance of the red marble being on vertex 3 and the blue marble being on vertex $b$. \\
\end{verse}

Now we may ask, how does a system with these {\em two} marbles change? What is its dynamics? Imagine that the
red marble is on vertex 1 and the blue marble is on vertex $a$. We may write this
as ``$1a$.'' What is the probability of the state going from state $1a$ to state $2b$? Obviously, the
red marble must go from vertex 1 to vertex 2 and (multiply) the blue marble must go
from vertex $a$ to vertex $b$. The probability is $\f{1}{3} \times \f{2}{3} = \f{2}{9}$.
In general, for a system to go from state $ij$ to a state $i'j'$ we must multiply the
probability of going from state $i$ to state $i'$ with the probability of going from state
$j$ to state $j'$.
$$\xymatrix{ij \ar[rrrr]^{M[i',i] \times N[j',j]} &&&& i'j'}.$$
For the changes of all the states, we have to do this for all the entries.
We are really giving the tensor product of two matrices\footnote{Formally, the tensor product of matrices is a function
$$ \otimes: \C^{m \times m} \times \C^{n \times n} \lra \C^{m \times m} \otimes \C^{n \times n}=\C^{mn \times mn}$$
and it is defined as
$$(A \otimes B)[j,k]=A[j/n,k/m]\times B[j \mbox{ MOD }n,k \mbox{ MOD }m].$$}:

$$ M\otimes N \q=  \bbordermatrix{ & 0 & 1 & 2\\
0& 0\left[ \begin{array}{cc}
\f{1}{3} & \f{2}{3} \\
\f{2}{3} & \f{1}{3} \end{array} \right] & \f{1}{6} \left[ \begin{array}{cc}
\f{1}{3} & \f{2}{3} \\
\f{2}{3} & \f{1}{3} \end{array} \right]& \f{5}{6} \left[ \begin{array}{cc}
\f{1}{3} & \f{2}{3} \\
\f{2}{3} & \f{1}{3} \end{array} \right]\\ \\
1 & \f{1}{3}\left[ \begin{array}{cc}
\f{1}{3} & \f{2}{3} \\
\f{2}{3} & \f{1}{3} \end{array} \right] & \f{1}{2} \left[ \begin{array}{cc}
\f{1}{3} & \f{2}{3} \\
\f{2}{3} & \f{1}{3} \end{array} \right]& \f{1}{6} \left[ \begin{array}{cc}
\f{1}{3} & \f{2}{3} \\
\f{2}{3} & \f{1}{3} \end{array} \right]\\ \\
2 & \f{2}{3} \left[ \begin{array}{cc}
\f{1}{3} & \f{2}{3} \\
\f{2}{3} & \f{1}{3} \end{array} \right]& \f{1}{3} \left[ \begin{array}{cc}
\f{1}{3} & \f{2}{3} \\
\f{2}{3} & \f{1}{3} \end{array} \right]& 0\left[ \begin{array}{cc}
\f{1}{3} & \f{2}{3} \\
\f{2}{3} & \f{1}{3} \end{array} \right] }.$$
$$\q=\bbordermatrix{ & 0a & 0b & 1a & 1b & 2a & 2b \\
0a& 0&0&\f{1}{18}&\f{2}{18}&\f{5}{18}&\f{10}{18}\\
0b& 0&0&\f{2}{18}&\f{1}{18}&\f{10}{18}&\f{5}{18}\\
1a& \f{1}{9}&\f{2}{9}&\f{1}{6}&\f{2}{6}&\f{1}{18}&\f{2}{18}\\
1b&\f{2}{9}&\f{1}{9}&\f{2}{6}&\f{1}{6}&\f{2}{18}&\f{1}{18}\\
2a&\f{2}{9}&\f{4}{9}&\f{1}{9}&\f{2}{9}&0&0\\
2b&\f{4}{9}&\f{2}{9}&\f{2}{9}&\f{1}{9}&0&0
}$$
The graph that corresponds to this matrix, $G_M\times G_N$ is called the Cartesian product of
two weighted graphs.
In quantum theory, the states of two systems are combined using the tensor product
of two vectors and the dynamics of two systems are combined by using the tensor
product of two matrices. The tensor product of the matrices will then act on the tensor
product of the vectors.

In general, the Cartesian product of an $n$-vertex graph with an $n'$-vertex graph is
an $(n\times n')$-vertex graph. If we have an $n$-vertex graph $G$ and we are interested in
$m$ different marbles moving in this system, we would need to look at the graph
$$G^m \q= \underbrace{G \times G \times \cdots \times G}_{m \mbox{ times}}$$ which will have
$n^m$ vertices. If $M_G$ is the associated adjacency matrix, then we will be interested in
$$M_G^{\otimes m} \q= \underbrace{M_G \otimes M_G \otimes \cdots \otimes M_G}_{m \mbox{ times}}$$
which will be a $n^m$ by $n^m$ matrix.

One might think of a bit as a two vertex graph with a marble on the $0$ vertex or a marble on the
$1$ vertex. If one were then interested in $m$ bits, one would need a $2^m$ vertex graph or equivalently
a $2^m$ by $2^m$ matrix. So there is exponential growth of the resources needed for the amount of bits
discussed.

This exponential growth for a system is actually one of the
main reasons why Richard Feynman started thinking about quantum computing.
He realized that because of this exponential growth, it would be hard for a classical
computer to simulate such a system. He asked whether a quantum computer with its ability to
do massive parallel processing, might be able to simulate such a system.

\section{Basic Quantum Theory}
Armed with the intuition, we tackle the formal statement of quantum theory. A disclaimer is
in order: We are only presenting a small part of finite dimensional quantum physics. There is no way that
we can give more than a minute fraction of this magnificent subject in these few pages. It is a sincere hope
that this will inspire the reader to go on and study more. For a mathematician, the best book to
start reading about quantum mechanics is, of course, Dirac's classic text \cite{Dirac}. However, there are
many other primers available, e.g., \cite{Chester, Martin, Polk} and \cite{White}. The more advanced
mathematician might want to look at \cite{Sudbery}.

\subsection{States}
An $n$ dimensional quantum system is a system that can be observed in one of $n$ possible states.
Examples of such systems are
\begin{itemize}
\item a particle can be in one of $n$ positions;
\item a system might have one
of $n$ energy levels;
\item a photon might have one of $n$ polarization directions.
\end{itemize}
For clarity, lets talk of the first example. Lets say we have a particle that can be in one of
$n$ positions.

The states of such a system shall be represented by collum vectors of $n$ complex numbers. We
shall denote these vectors with the ``ket'' $|\quad \ra$ \footnote{``Ket'' is the second half of ``bracket''.
However we shall not use the ``bra'' part in our exposition.}
 notation:
$$|\varphi\ra = [c_0, c_1, \ldots, c_j, \ldots, c_{n-1}]^T.$$
How is one to interpret these kets? Let us look at simple cases. The state
$$|\psi\ra=[0,1, \ldots, 0, \ldots, 0]^T$$
is to be thought of as saying that our particle will be found in position 1. The
state
$$|\psi'\ra=[0,0, \ldots, 1, \ldots, 0]^T$$ is to be interpreted that the
particle is in position $j$. These two states are examples of what are
called ``pure states''.

How is one to interpret an arbitrary $$|\varphi\ra = [c_0, c_1, \ldots, c_j, \ldots, c_{n-1}]^T?$$
Let $S$ be the sum of the squares of modulus of the $c_j$, i.e.,
$$S=|c_0|^2+|c_1|^2+\ldots+|c_{n-1}|^2.$$
This is the length of the vector $|\varphi\ra $. Then $|\varphi\ra$ is to be interpreted that if one
was to measure the state described by $|\varphi\ra$ we would find the particle in position
$0$ with probability $|c_0|^2/S$, in position $1$ with probability $|c_1|^2/S$, in position
$2$ with probability $|c_2|^2/S$, $\ldots$, in position $n-1$ with probability $|c_{n-1}|^2/S$
Such states are called ``superpositions''. They say that the particle is in more than one
``position'' at a time. It is important to stress that $|\varphi\ra$
means that the particle is in {\em all} positions simultaneously. It does not
mean that the particle is in some single position and the $c_j$ are giving
us probabilities of which position.

These superpositions can be added:
if
$$|\varphi\ra = [c_0, c_1, \ldots, c_j, \ldots, c_{n-1}]^T$$
and
$$|\varphi'\ra = [c'_0, c'_1, \ldots, c'_j, \ldots, c'_{n-1}]^T,$$
then
$$|\varphi\ra + |\varphi'\ra= [c_0+c'_0, c_1+c'_1, \ldots, c_j+c'_j, \ldots, c_{n-1}+c'_{n-1}]^T.$$
Also, if there is a complex number $c\in \C$, we can multiply a ket by this $c$:
$$c |\varphi\ra = [c\times c_0, c\times c_1, \ldots, c\times c_j, \ldots, c\times c_{n-1}]^T$$
These operation satisfy all the properties of being a complex vector space.
So the states of an $n$ dimensional quantum system are represented by the complex
vector space $\C^n$.

Let us add a superposition to itself.
$$|\varphi \ra +|\varphi \ra= 2|\varphi\ra = [c_0+c_0, c_1+c_1, \ldots, c_j+c_j, \ldots, c_{n-1}+c_{n-1}]^T$$
$$=[2c_0, 2c_1, \ldots, 2c_j, \ldots, 2c_{n-1}]^T.$$
The sum of the modulus squared is
$$S'=|2c_0|^2+|2c_1|^2+\ldots+|2c_{n-1}|^2=2^2|c_0|^2+2^2|c_1|^2+\ldots+2^2|c_{n-1}|^2=2^2(|c_0|^2+|c_1|^2+\ldots+|c_{n-1}|^2).$$
The chances of a particle being found in position $j$ is
$$\f{|2c_j|^2}{S'}=\f{2^2|c_j|^2}{2^2(|c_0|^2+|c_1|^2+\ldots+|c_{n-1}|^2)}=\f{|c_j|^2}{|c_0|^2+|c_1|^2+\ldots+|c_{n-1}|^2}.$$
In other words, the ket $2|\varphi \ra$ describes the same physical system as $|\varphi \ra.$ This makes sense, after all
when we add two of the same superpositions, we do not expect any interference. We expect that they should
reinforce each other. A similar analysis shows that for an arbitrary $c \in \C$ we have that the ket $|\varphi \ra$ and the
ket $c|\varphi \ra$ describe the same physical state. Geometrically, we can say that the vector $|\varphi \ra$ and the extension
$c|\varphi \ra$ describe the same physical state. So the only thing that is important is the direction of $|\varphi \ra$ not
the length of $|\varphi \ra$. We might as well work with a ``normalized'' $|\varphi \ra$, i.e.,
$$\f{|\varphi \ra}{||\varphi \ra|}$$ which has length 1. In fact, in section 1, we only worked with normalized vectors.

Given an $n$ dimensional quantum system represented by $\C^n$ and an $m$ dimensional quantum system represented by$\C^m$, we can combine these two systems
to form one system. This one system is to represented by the tensor product of the two vector spaces:
$$\C^n \otimes \C^m \cong \C^{n \times m}.$$ If $|\varphi\ra$ is in the first system
and $|\varphi'\ra $ is in the second system, then we represent the
combined system as
$$|\varphi\ra \otimes |\varphi'\ra = |\varphi, \varphi'\ra = |\varphi \varphi' \ra.$$
It is important to realize that, in general, there are more elements in the tensor
product of the two systems than in the union of each of the two systems. States in $\C^n \otimes \C^m$ that cannot be represented
simply as an element in $\C^n$ and an element in $\C^m$ are said to be ``entangled''.

\subsection{Dynamics}
Quantum systems are not static. The states of the system are constantly changing. Changes, or ``operators'', on an $n$ dimensional
quantum system
are represented by $n$ by $n$ unitary matrices. Given a state $|\varphi\ra$ that represents a system at time $t$, then
the system will be in state
$U |\varphi\ra$
at time $t+1$.

What does unitary really mean? If $U|\varphi\ra=|\varphi'\ra$ then we can easily form $U^\dag$ and multiply both sides of the equation
by $U^\dag$ to
get $U^\dag U|\varphi\ra =U^\dag |\varphi'\ra$ or $|\varphi\ra=U^\dag |\varphi'\ra$. In other words, because $U$ is unitary, there is a related matrix
that can ``undo'' the action that $U$ does. $U^\dag$ takes the result of $U$'s action and gets the original
vector back. In the quantum world, most actions actions are ``undoable'' or ``reversible'' in such a manner.

If $U$ operates on $\C^n$ and $U'$ operates on $\C^m$, then $U \otimes U'$ will operate on $\C^n \otimes \C^m$ in the following way:
$$(U \otimes U')(|\varphi\ra \otimes |\varphi'\ra)= U|\varphi\ra \otimes U'|\varphi'\ra.$$
\subsection{Observables}
There are other types of operations that one can do to a $n$ dimensional quantum system: one can observe, or ``measure'', the system.
When we measure a system, it is no longer in a superposition. The superposition is said to ``collapse'' to a pure state.
$$|\varphi\ra = [c_0, c_1, \ldots, c_j, \ldots, c_{n-1}]^T \rightsquigarrow |\varphi'\ra = [0, 0, \ldots, 1, \ldots, 0].^T$$

The question is which of the $n$ pure states will the state collapse to? The answer is that it is
random. Let $S$ be the sum of all the squares of the modulus, i.e.,
$$S=|c_0|^2+|c_1|^2+\cdots + |c_j|^2+ \cdots +|c_{n-1}|^2. $$
There is a $|c_0|^2/S$ of a chance of the superposition collapsing to the 0th pure state.
There is a $|c_1|^2/S$ of a chance of the superposition collapsing to the 1th pure state. etc.
There is no known mathematical way to decide which pure state the system will, in-fact, collapse to.

An observable, or ``measurement'' on an $n$ dimensional system is represented by an $n$ by $n$ hermitian matrix.
We remind the reader that a $n$ by $n$ matrix $A$ is hermitian, or ``self-adjoint''
if $A^\dag=A$. In other words,
$A[j,k] = \overline{A[k,j]}$. Equivalently $A$
is hermitian if and only if $A^T=\overline{A}$.
For example, the matrix
$$\left[ \begin{array}{ccc}
5 & 4+5i & 6-16i \\
4-5i & 13 & 7 \\
6+16i & 7 & -2.1
\end{array} \right] $$ is hermitian.

For a matrix $A$ in $\C^{n \times n}$, if there is a number $c$ in $\C$ and a vector $|\psi\ra$ in $\C^n$ such that
$$A|\psi\ra  =  c|\psi\ra$$
then $c$ is called an ``eigenvalue'' of $A$ and $|\psi\ra $ is called an ``eigenvector'' of $A$ associated to $c$.
The eigenvalues of a hermitian matrix are all real numbers. Furthermore, distinct
eigenvectors which have distinct eigenvalues of any hermitian matrix are orthogonal.
The hermitian matrices that represent observables for an $n$ dimensional system have the further property that there are
$n$ distinct eigenvalues and $n$ distinct eigenvectors. That means that the set of eigenvectors form a basis for the
entire complex vector space that represents the quantum system we are interested in. Hence if we have an observable $A$
and $|\varphi\ra$ an eigenvalue of $A$ then $A|\varphi\ra =c|\varphi\ra$ for some $c \in \C$. $c|\varphi\ra$ represents the
same state as $|\varphi\ra$ as we said before. So if the system is in an eigenvector of the basis, then the
system will not change.

\section{Architecture}
In this section we are going to show how the ideas of quantum mechanics are
going to influence our construction of quantum computers. In this paper we
do not dwell on actual hardware implementations. Rather we shall look at
the quantum generalizations of bits and logical gates. In section 3.1 we go from bits
to qubits. We also discuss the notation that is needed for this. In section 3.2 we
show how to look at classical computing as matrix manipulations. From the view afforded
by this perspective, we easily generalize the notion of logical gate to quantum gate.
There are many quantum gates, but we shall only look at a few that will be needed in
the next section.

\subsection{Bits and Qubits}
What is a bit? A bit is an atom of information that represents one of two
disjoint situations. An example of a bit is electricity going through a circuit or
electricity not going through a circuit; a switch turned on
or a switch turned off; a way of saying true or false. All
these examples are saying the same thing: a bit is a way of
describing a system whose set of states is of size two.

A bit can be represented by two 2 by 1 matrices. We shall represent 0---or better the state
$|0\ra$ as
$$ |0\ra \q=\bbordermatrix{
& \\
0 & 1 \\
1& 0} $$ We shall represent a 1, or state $|1\ra$ as:
$$ |1\ra \q= \bbordermatrix{
& \\
0 & 0 \\
1& 1} $$ Since these are two different representations, we have an honest-to-goodness bit.

A bit is either in state $|0\ra$ or in state $|1\ra$. That was
sufficient for the classical world.  But that is not
sufficient for the quantum world. In that world we have situations
where we are in one state {\it and} in the other state
simultaneously. In the quantum world we have systems where a switch
is in a superposition of states on {\it and } off. So we define a ``quantum bit'' or a
``qubit'' as a way of describing a quantum system of dimension two.  We shall represent any such qubit as a two by one matrix with complex numbers:
$$\bbordermatrix{
& \\
0 & c_0 \\
1& c_1} $$ where $|c_0|^2+|c_1|^2=1$. Notice that a classical bit is
a special type of qubit. $|c_0|^2$ is to be interpreted as the
probability that after measuring the qubit, it will be found in
state $|0\ra$. $|c_1|^2$ is to be interpreted as the probability
that after measuring the qubit it will be found in state $|1\ra$.
Whenever we measure a qubit, it automatically becomes a bit. So we
shall never ``see'' a general qubit. Nevertheless, they do exist and
they are the core of our tale. The power of quantum computers is due
to the fact that a system can be in many states at the same time.

Following the normalization procedure that we learned in the last section, any
element of $\C^2$ can be converted into a qubit. For example, the
vector $$V=\left[\begin{matrix}5+3i\\6i \end{matrix}\right]$$ has
norm $$|V| = \sqrt{\la V,V\ra} = \sqrt{\left[ 5-3i, -6i\right]
\left[\begin{matrix}5+3i\\6i
\end{matrix}\right]}=\sqrt{34+36}=\sqrt{70}.$$ So $V$ describes the
same physical state as the qubit
$$\f{V}{\sqrt{70}}=\left[\begin{matrix}\f{5+3i}{\sqrt{70}}\\\f{6i}{\sqrt{70}}
\end{matrix}\right].$$ After measuring the qubit $\f{V}{\sqrt{70}}$,
the probability of being in state $|0\ra$ is $\f{34}{70}$, and the
probability of being in state $|1\ra$ is $\f{36}{70}$.

It is easy to see that the bits $|0\ra$ and $|1\ra$ are the
canonical basis of $\C^2$. So any qubit can be written as
$$\left[\begin{matrix}c_0\\c_1\end{matrix}\right]\q= c_0 \cdot \left[\begin{matrix}1\\0 \end{matrix}\right] +
c_1 \cdot \left[\begin{matrix}0\\1\end{matrix}\right] \q= c_0|0\ra
+c_1|1\ra.$$

Let us look at several ways of writing different qubits.
$\f{1}{\sqrt{2}}\left[\begin{matrix}1\\1 \end{matrix}\right]$ can be
written as
$$\left[\begin{matrix}\f{1}{\sqrt{2}}\\ \f{1}{\sqrt{2}} \end{matrix}\right] =
\f{1}{\sqrt{2}}|0\ra + \f{1}{\sqrt{2}}|1\ra =
\f{|0\ra+|1\ra}{\sqrt{2}}.$$ Similarly
$\f{1}{\sqrt{2}}\left[\begin{matrix}1\\-1 \end{matrix}\right]$ can
be written as
$$\left[\begin{matrix}\f{1}{\sqrt{2}}\\ \f{-1}{\sqrt{2}} \end{matrix}\right]
= \f{1}{\sqrt{2}}|0\ra - \f{1}{\sqrt{2}}|1\ra = \f{|0\ra -
|1\ra}{\sqrt{2}}.$$

It is important to realize that
$$\f{|0\ra+|1\ra}{\sqrt{2}}=\f{|1\ra+|0\ra}{\sqrt{2}}.$$ These are
both ways of writing
$\vw{\f{1}{\sqrt{2}}}{\f{1}{\sqrt{2}}}$.
 In contrast,
$$\f{|0\ra-|1\ra}{\sqrt{2}}\ne \f{|1\ra-|0\ra}{\sqrt{2}}$$
The first state is $\vw{\f{1}{\sqrt{2}}}{-\f{1}{\sqrt{2}}}$ and the
second one is $\vw{-\f{1}{\sqrt{2}}}{\f{1}{\sqrt{2}}}$. However the
two states are related:
$$\f{|0\ra-|1\ra}{\sqrt{2}}= (-1)\f{|1\ra-|0\ra}{\sqrt{2}}.$$

How are qubits to be implemented?  While this is not our focus, some examples of
qubit implementations are given:
\begin{itemize}
\item An electron might be in one of two different orbits around a nucleus of an atom. (Ground state and excited state.)
\item A photon might be in one of two different polarized states.
\item A subatomic particle might be in spinning in one of two different directions.
\end{itemize}
There will be enough quantum indeterminacy and quantum superposition
effects within all these systems to represent a qubit.

\vspace{.5in} Computers with only one bit of storage are not very interesting.
Similarly, we will need quantum devices with more then one qubit. Consider a
byte, or eight bits. A typical byte might look like
$$01101011.$$
We might also write it as
$$
\left[ \begin{matrix}
1 \\
0\end{matrix}\right] , \left[ \begin{matrix}
0 \\
1\end{matrix}\right] , \left[ \begin{matrix}
0 \\
1\end{matrix}\right] , \left[ \begin{matrix}
 1 \\
0\end{matrix}\right] , \left[ \begin{matrix}
 0 \\
 1\end{matrix}\right] ,
\left[ \begin{matrix}
 1 \\
 0\end{matrix}\right] ,
\left[ \begin{matrix}
 0 \\
1\end{matrix}\right] , \left[ \begin{matrix}
0 \\
1\end{matrix}\right].$$ Previously, we learned that in order to
combine systems one should use the tensor product. We can describe
the above byte as
$$|0\ra \otimes |1\ra \otimes |1\ra \otimes |0\ra \otimes |1\ra \otimes |0\ra \otimes |1\ra \otimes |1\ra .$$
As a qubit, this is an element of
$$\C^2 \otimes \C^2 \otimes\C^2 \otimes\C^2 \otimes\C^2 \otimes\C^2 \otimes\C^2 \otimes\C^2. $$
This vector space can be written as
$(\C^2)^{\otimes 8}$. This is a complex vector space of dimension
$2^8=256$. Since there is only one complex vector space (up to
isomorphism) of this dimension, this space is isomorphic to
$\C^{256}$.

Our byte can be described as in yet another way:
As a $2^8=256$ row vector
$$\bbordermatrix{
& \\
00000000 & 0 \\
00000001 & 0 \\
\vdots & \vdots \\
01101010 & 0\\
01101011 & 1\\
01101100 & 0\\
\vdots & \vdots \\
11111110 & 0\\
11111111 & 0 }.$$

This is fine for the classical world. However, for the quantum world,
a general qubit can be written as
$$\bbordermatrix{
& \\
00000000 & c_0 \\
00000001 & c_1 \\
\vdots & \vdots \\
01101010 & c_{106}\\
01101011 & c_{107}\\
01101100 & c_{108}\\
\vdots & \vdots \\
11111110 & c_{254}\\
11111111 & c_{255} }$$ where $\sum_{i=0}^{255} |c_i|^2 =1$.

In the classical world, you need to write the state of each of the
eight bits. This amounts to writing eight bits. In the quantum world,
a state of eight qubits is given by writing 256 complex numbers. This exponential growth was one of the
reasons why researchers started thinking about quantum computing. If
you wanted to emulate a quantum computer with a 64 qubit register, you
would need to store $2^{64} = 18,446,744,073,709,551,616$ complex
numbers. This is beyond our current ability.

Let us practice writing two qubits in ket notation. The qubits
$$\bbordermatrix{
 & \\
 00 & 0 \\
 01 & 1 \\
 10 & 0 \\
 11 & 0 }$$ can be written as $$|0\ra \otimes |1\ra.$$ Since the tensor
 product is understood, we might also write these qubits as
 $|0,1\ra$ or $|01\ra$. The qubit corresponding to $$\f{1}{\sqrt{3}}\left[\begin{matrix}1\\0\\-1\\1\end{matrix}\right]$$
 can be written as $$\f{1}{\sqrt{3}}|00\ra - \f{1}{\sqrt{3}}|10\ra + \f{1}{\sqrt{3}} |11\ra = \f{|00\ra - |10\ra + |11\ra}{
 \sqrt{3}}.$$

The tensor product of two states is not commutative.
$$|0\ra \otimes |1\ra=|0,1\ra=|01\ra \ne |10\ra =|1,0\ra =|1\ra \otimes |0\ra. $$
The first ket says that the first qubit is in state 0 and the second
qubit is in state 1. The second ket says that first qubit is in
state 1 and the second state is in state 0.

\subsection{Classical Gates}
Classical logical gates are ways of manipulating bits. Bits go into
logical gates and bits come out. We
represent $n$ input bits as a $2^n$ by 1 matrix and $m$ output bits
as a $2^m$ by 1 matrix. How should we represent our logical gates? A
$2^m$ by $2^n$ matrix takes a $2^n$ by 1 matrix and outputs a $2^m$
by 1 matrix. In symbols:
$$(2^m \times 2^n)(2^n \times 1) = (2^m \times 1).$$
So bits will be represented by column vectors and logic gates will
be represented by matrices.

Let us do a simple example. Consider the NOT gate.
NOT takes as input one bit, or a $2$ by 1 matrix,  and outputs one
bit, or a 2 by 1 matrix. NOT of $|0\ra$ equals $|1\ra$ and NOT of
$|1\ra$ equals $|0\ra$. The matrix
$$\mbox{NOT} = \left[ \begin{matrix} 0&1\\1&0 \end{matrix} \right].$$ This matrix satisfies
$$\left[ \begin{matrix} 0&1\\1&0 \end{matrix} \right]\left[ \begin{matrix} 1\\0 \end{matrix} \right]=
\left[ \begin{matrix} 0\\1 \end{matrix} \right] \qquad \left[
\begin{matrix} 0&1\\1&0 \end{matrix} \right]\left[ \begin{matrix}
0\\1 \end{matrix} \right]= \left[ \begin{matrix} 1\\0 \end{matrix}
\right],$$ which is exactly what we want.

What about the other gates? The other gates will be given by the following matrices:
$$\begin{array}{||c|c|c|c||}\hline \hline AND&NAND&OR&NOR\\ \hline
\left[ \begin{matrix}
1&1&1&0\\0&0&0&1 \end{matrix} \right]&\left[ \begin{matrix}0&0&0&1\\1&1&1&0\end{matrix} \right]
&\left[ \begin{matrix}
1&0&0&0\\0&1&1&1 \end{matrix} \right]&\left[ \begin{matrix}
0&1&1&1\\1&0&0&0 \end{matrix} \right]\\ \hline \hline
\end{array}$$

When we perform a computation, often we have to carry out one
operation followed by another.

\vspace{.5in}

$$\Qcircuit
@C=1em @R=5em
{&\qw&\qw&\gate{\quad \begin{array}{c} A  \\ \end{array} \quad}&\qw&\gate{\quad \begin{array}{c}B \\ \end{array} \quad}&\qw&\qw }$$

\vspace{.5in}

We shall call this performing ``sequential'' operations. If matrix
$A$ corresponds to performing an operation and matrix $B$
corresponds to performing another operation, then the multiplication
matrix $B \star A$ corresponds to performing the operation
sequentially.

Besides sequential operations, there are ``parallel'' operations:
\vspace{.5in}
$$\Qcircuit
@C=1em @R=1em
{&\qw&\qw&\gate{\quad \begin{array}{c}A  \\ \end{array} \quad}&\qw&\qw&\qw \\
&\qw&\qw&\gate{\quad \begin{array}{c} B  \\ \end{array} \quad}&\qw&\qw&\qw
}$$
\vspace{.5in}
Here we are doing $A$ to some bits and $B$ to other bits. This will
be represented by $A \otimes B$, the tensor product of two matrices.

Combination of sequential and parallel operations gates/matrices
will be circuits.
A starting point is the realization that
the circuit

\vspace{.5in}

$$\Qcircuit
@C=1em @R=1em
{&\qw&\qw&\gate{\quad \begin{array}{c} A  \\ \end{array} \quad}&\qw&\gate{\quad \begin{array}{c} B  \\ \end{array} \quad}&\qw&\qw\\
&\qw&\qw&\gate{\quad \begin{array}{c} A' \\ \end{array} \quad}&\qw&\gate{\quad \begin{array}{c} B'  \\ \end{array} \quad}&\qw&\qw
}$$
\vspace{.5in}
can be realized as
$$(B \star A) \otimes (B' \star A')=(B \otimes B')\star(A \otimes A').$$
We will of course have
some really complicated matrices, but they will all be decomposable
into the sequential and parallel composition of simple gates.
\subsection{Quantum Gates}
A quantum gate is simply any unitary matrix that manipulates qubits. There are some simple gates that are
quantum gates.

The ``Hadamard matrix''
$$ H = \f{1}{\sqrt{2}}\left[ \begin{array}{cc}1&1\\1&-1\end{array}\right] = \left[ \begin{array}{cc}
\f{1}{\sqrt{2}}&\f{1}{\sqrt{2}}\\
\f{1}{\sqrt{2}}&-\f{1}{\sqrt{2}}
\end{array} \right].$$
The Hadamard matrix is also its own inverse.
As it turns out, the Hadamard matrix is one of the most important matrices in quantum computing.

Consider the following ``controlled-not'' gate.

\vspace{.51in}

$$\Qcircuit
{&\ustick{|x\rangle}\qw&\ctrl{1}&\ustick{|x\rangle}\qw&\qw\\
&\ustick{|y\rangle}\qw&\targ&\ustick{|( x\oplus y)\rangle}\qw&\qw }$$

\vspace{.51in}

This gate has two inputs and has two outputs. The top input is the control bit. It controls what the
output will be. If $|x\ra=|0\ra$, then the output of $|y\ra$ will be the same as the input. If $|x\ra=|1\ra$
then the output of $|y\ra$ will be the opposite. If we write the top qubit first and then the bottom qubit, then
the controlled-not gate takes $|x,y\ra$ to $|x,x\oplus y\ra $ where $\oplus$ is the binary exclusive or operation.

The matrix that corresponds to this reversible gate is
$$\bbordermatrix {
& 00&01&10& 11 \\
00 & 1&0&0&0   \\
01 & 0&1&0&0   \\
10 & 0&0&0&1   \\
11 & 0&0&1&0}.$$

One last piece of notation: we shall denote an observation (measurement) by the following ``meter'':

\vspace{.51in}

$$\Qcircuit
{&\meter}$$

\vspace{.51in}

There are many other quantum gates, but we shall not need them for our work in the next section.

\section{Deutsch's Algorithm}
The simplest quantum algorithm is Deutsch's algorithm which is a
nice short algorithm that solves a slightly contrived problem. This
algorithm is concerned with functions from the set $\{0,1\}$ to the
set $\{0,1\}$. There are four such functions which we might
visualize as
$$\xymatrix{
0 \bullet \ar@{|->}[r]& \bullet 0&0 \bullet \ar@{|->}[r] & \bullet 0&0 \bullet \ar@{|->}[rd]& \bullet 0&0 \bullet \ar@{|->}[rd]& \bullet 0\\
1 \bullet \ar@{|->}[ru]& \bullet 1&1 \bullet  \ar@{|->}[r]& \bullet 1&1 \bullet
\ar@{|->}[ru]& \bullet 1&1 \bullet \ar@{|->}[r] & \bullet 1& }$$

Call a function $f:\{0,1\} \lra \{0,1\}$, ``balanced''  if
$f(0)\neq f(1))$. In contrast, call a function ``constant''  if
$f(0) = f(1)$. Of the four functions, two are balanced and two are
constant.

Deutsch's algorithm solves the following problem: Given a function
$f:\{0,1\} \lra \{0,1\}$ as a black-box, where one can evaluate an
input, but cannot ``look inside'' and ``see'' how the function is
defined, tell if the function is balanced or constant.

With a classical computer, one would have to first evaluate $f$ on
an input, then evaluate $f$ on the second input and then compare the
outputs.
The point is that with a classical computer, $f$ must be evaluated
twice. Can we do better?

A quantum computer can be in two states at one time. We shall use
this superposition of states to evaluate both inputs at one time.

\vspace{.5in} In classical computing, evaluating a given function
$f$ would correspond to performing the following operation

$$\Qcircuit
@C=1em @R=5em
{&\ustick{x}\qw&\qw&\gate{\qquad \begin{array}{c}\\ f \\ \\ \end{array} \qquad}&\qw&\ustick{f(x)}\qw&\qw }$$

Such a function could be
thought of as a matrix. The function
$$\xymatrix{
0 \bullet \ar@{|->}[r]& \bullet 0\\
1 \bullet \ar@{|->}[ru]& \bullet 1 }$$ is equivalent to the matrix
$$\bbordermatrix{& 0&1 \\ 0 & 1&1 \\ 1 & 0&0 }.$$ Multiplying either
state $|0\ra$ or state $|1\ra$ on the right of this matrix would
result in state $|0\ra$. The column name is to be thought of as the
input and the row name is to be thought of as the output.

However, this will not be good for a quantum system. For a quantum
system we need a little something extra. A quantum system must be
unitary (reversible). Given the output, we must be able to find the
input. If $f$ is the name of the function, then the following
black-box $U_f$ system will be the quantum gate that we shall employ
to evaluate input:

\vspace{.5in}

$$\Qcircuit
{&\ustick{|x\rangle}\qw&\qw&\multigate{1}{\qquad U_f \quad }&\qw&\ustick{|x \rangle}\qw&\qw\\
&\ustick{|y\rangle}\qw&\qw&\ghost{\qquad U_f
\quad}&\qw&\ustick{|y\oplus f(x)\rangle}\qw&\qw. }$$

\vspace{.5in}

The top input, $|x\ra$, will be the qubit value that one wishes to
evaluate and the bottom input, $|y\ra$, controls the output. The top
output will be the same as the input qubit $|x\ra$ and the bottom
output will be the qubit $|y \oplus f(x)\ra$ where $\oplus$ is XOR,
the exclusive or operation. We are going to
write from left to right the top qubit first and then the bottom. So
we  say that this function takes the state $|x,y\ra$ to the state
$|x,y\oplus f(x)\ra$. If $y=0$ this simplifies: $|x,0\ra$ to
$|x,0\oplus f(x)\ra=|x,f(x)\ra.$ This gate is reversible by simply
looking at the following system

\vspace{.5in}

$$\Qcircuit
{&\ustick{|x\rangle}\qw&\multigate{1}{\qquad U_f \quad }&\qw&\ustick{|x \rangle}\qw&
\qw&\multigate{1}{\qquad U_f \quad }&\qw&\ustick{|x \rangle}\qw&\qw\\
&\ustick{|y\rangle}\qw&\ghost{\qquad U_f
\quad}&\qw&\ustick{|y \oplus f(x)\rangle}\qw&\qw\qw&\ghost{\qquad U_f
\quad}&\qw&\ustick{| y\rangle}\qw&\qw }$$

\vspace{.5in}

\noindent State $|x,y\ra$ goes to $|x,y\oplus f(x)\ra$ which further
goes to
$$|x,(y \oplus f(x))\oplus f(x)\ra=|x,y\oplus (f(x) \oplus f(x))\ra= |x, y \oplus 0\ra = |x,y \ra.$$

In quantum systems, evaluating $f$ is equivalent to multiplying a
state of the input by a unitary matrix $U_f$. For the function $$\xymatrix{
0 \bullet \ar@{|->}[rd]& \bullet 0\\
1 \bullet \ar@{|->}[ru]& \bullet 1 }$$ the corresponding unitary matrix is
$$\bbordermatrix{ &00&01&10&11
\\
00&0&1&0&0\\
01&1&0&0&0\\
10&0&0&1&0\\
11&0&0&0&1 }.$$

Remember that the top column name correspond to the input $|x,y\ra$
and the left-hand row name corresponds to the outputs $|x, y \oplus
f(x) \ra$. A $1$ in the $xy$ column and the $x'y'$ row means for
input $|x,y\ra$ the output will be $|x',y'\ra$.

So we are given such a matrix that expresses a function but we cannot ``look inside'' the matrix to ``see'' how it is defined. We are
asked to determine if the function is balanced or constant.
\subsection{First Attempt}
Let us take a first stab at a quantum algorithm to
solve this problem. Rather than evaluating $f$ twice, let us try our
trick of superposition of states. Instead of having the top input to
be either in state $|0\ra$ or in state $|1\ra$, we shall put the top
input in state $$\f{|0\rangle + |1\rangle}{\sqrt{2}}$$ which is
``half-way'' $|0\ra$ and ``half-way'' $|1\ra$. The Hadamard matrix can place a qubit in such a state.
$$ H |0\ra = \left[ \begin{array}{cc}
\f{1}{\sqrt{2}}&\f{1}{\sqrt{2}}\\
\f{1}{\sqrt{2}}&-\f{1}{\sqrt{2}}
\end{array} \right]\left[ \begin{array}{c}
1\\0\end{array}
\right]=\vw{\f{1}{\sqrt{2}}}{\f{1}{\sqrt{2}}}=\f{|0\rangle +
|1\rangle}{\sqrt{2}}.$$ The obvious (and not necessarily
correct) state to put the bottom input is as state $|0\ra$. And so
we have:

\vspace{.5in}

$$\Qcircuit
{&\ustick{|0\rangle}\qw&\qw&\gate{H}&\qw&\multigate{1}{\qquad U_f \quad }&\qw&\qw&\qw&\meter\\
&\ustick{|0\rangle}\qw&\qw&\qw&\qw&\ghost{\qquad U_f \quad }&\qw&\qw&\qw&\qw\\
&&\Uparrow && \Uparrow &&& \Uparrow \\
&&|\varphi_0\rangle && |\varphi_1\rangle &&& |\varphi_2\rangle }$$

\vspace{.5in}

In terms of matrices this corresponds to
$$ U_f (H \otimes I)(|0\ra \otimes|0\rangle ).$$

We shall carefully examine the states of the system at every point.
The system starts in
$$|\varphi_0 \rangle \q= (|0\ra \otimes|0\rangle )\q= |0,0\rangle .$$
We then apply the Hadamard matrix only to the top input --- leaving
the bottom input alone --- to get
$$|\varphi_1 \rangle \q= \left[ \f{|0\rangle + |1\rangle}{\sqrt{2}}\right] |0\rangle  \q=
\f{|0,0\rangle +|1,0\rangle}{\sqrt{2}}.$$ After multiplying with
$U_f$ we have
$$ |\varphi_2 \rangle \q= \f{|0,f(0)\rangle + |1,f(1)\rangle}{\sqrt{2}}$$

For the function $0\mapsto 1$ and $1 \mapsto 0$ the state $ |\varphi_2 \rangle$
would be
$$ |\varphi_2 \rangle=\bbordermatrix{ &00&01&10&11
\\
00&0&1&0&0\\
01&1&0&0&0\\
10&0&0&1&0\\
11&0&0&0&1
}\bbordermatrix{&\\
00&\f{1}{\sqrt{2}}\\
01 &0\\
10 &\f{1}{\sqrt{2}}\\
11&0}=\bbordermatrix{&\\
00&0\\
01 &\f{1}{\sqrt{2}}\\
10 &\f{1}{\sqrt{2}}\\
11&0}=\f{|0,1\rangle + |1,0\rangle}{\sqrt{2}}$$

If we measure the top qubit, there will be a 50-50 chance of finding
it in state $|0\ra$ and a 50-50 chance of finding it in state $|1\ra$. Similarly there is no real
information to be gotten by measuring the bottom qubit. So the
obvious algorithm does not work. We need another trick.

\vspace{.5in}
\subsection{Second Attempt}
Let us take another stab at solving our problem. Rather then leaving
the bottom qubit in state $|0\ra$, let us put it in the
superposition state:
 $$\f{|0\rangle - |1\rangle}{\sqrt{2}}= \vw{\f{1}{\sqrt{2}}}{- \f{1}{\sqrt{2}}}. $$
Notice the minus sign. Even though there is a negation, this state
is also ``half-way'' in state $|0\ra$ and ``half-way'' in state
$|1\ra$. The change of phase will help us get our desired results.
We can get to this superposition of states by multiplying state
$|1\ra$ with the Hadamard matrix. We shall leave the top qubit as an
ambiguous $|x\ra$.

\vspace{.5in}

$$\Qcircuit
{&\ustick{|x\rangle}\qw&\qw&\qw&\qw&\multigate{1}{\qquad U_f \quad }&\qw&\qw&\qw&\meter\\
&\ustick{|1\rangle}\qw&\qw&\gate{H}&\qw&\ghost{\qquad U_f \quad }&\qw&\qw&\qw&\qw\\
&&\Uparrow && \Uparrow &&& \Uparrow \\
&&|\varphi_0\rangle && |\varphi_1\rangle &&& |\varphi_2\rangle }$$

\vspace{.5in}

In terms of matrices, this becomes:
$$ U_f (I \otimes H)|x,1\rangle.$$

Let us look carefully at how the states of the qubits change.
$$|\varphi_0 \rangle \q= |x,1\rangle. $$
After the Hadamard matrix, we have
$$|\varphi_1 \rangle \q= |x \ra \left[ \f{|0\rangle - |1\rangle}{\sqrt{2}} \right]
\q= \f{|x,0\rangle - |x,1\rangle}{\sqrt{2}}.$$ Applying $U_f$ we get
$$|\varphi_2 \rangle \q= |x\ra \left[ \f{|0 \oplus f(x) \ra - |1 \oplus f(x) \ra}{\sqrt{2}} \right]
\q= |x\ra \left[\f{|f(x)\rangle -
|\overline{f(x)}\rangle}{\sqrt{2}}\right]$$ where $\overline{f(x)}$
means the opposite of $f(x)$. And so we have
$$|\varphi_2 \rangle =
 \left\{ \begin{array}{r @{\qquad}l}
|x\ra \left[ \f{|0 \ra - |1 \ra}{\sqrt{2}} \right] & \mbox{ if }f(x)=0 \\ \\
|x\ra \left[ \f{|1 \ra - |0 \ra}{\sqrt{2}} \right] & \mbox{ if
}f(x)=1 \end{array}\right. .$$ Remembering that $a-b=(-1)(b-a)$ we
might write this as
$$|\varphi_2 \rangle = (-1)^{f(x)} |x\ra
\left[ \f{|0 \ra - |1 \ra}{\sqrt{2}} \right].\label{evaluateatx}$$

What would happen if we evaluate either the top or the bottom state?
Again, this does not really help us. We do not gain any information. The top qubit will
be in state $|x\ra$ and the bottom qubit will --- with equal probability --- be either in state
$|0\ra$ or in state $|1\ra$. We need something more.

\subsection{Deutsch's Algorithm}
Now let us combine both of these attempts to actually give Deutsch's
algorithm.
Deutsch's algorithm works by putting {\it both} the top and bottom
qubits into a superposition. We will also put the results of the top
qubit through a Hadamard matrix.

\vspace{.5in}

$$\Qcircuit
{&\ustick{|0\rangle}\qw&\qw&\gate{H}&\qw&\multigate{1}{\qquad U_f \quad }&\qw&\gate{H}&\qw&\meter\\
&\ustick{|1\rangle}\qw&\qw&\gate{H}&\qw&\ghost{\qquad U_f \quad }&\qw&\qw&\qw&\qw\\
&&\Uparrow&&\Uparrow&&\Uparrow&&\Uparrow \\
&&|\varphi_0\rangle && |\varphi_1\rangle && |\varphi_2\rangle
&&|\varphi_3\rangle }$$

\vspace{.5in}

In terms of matrices this becomes:
$$ (I \otimes H) U_f (H \otimes H)|0,1\rangle$$

At each point of the algorithm the states are as follows.
$$|\varphi_0 \rangle = |0,1\rangle. $$
$$|\varphi_1 \rangle = \left[ \f{|0\rangle + |1\rangle}{\sqrt{2}} \right]
\left[\f{|0\rangle - |1\rangle}{\sqrt{2}}\right]\q=
\f{+|0,0\rangle - |0,1\rangle +|1,0\rangle -|1,1\rangle}{2}=\bbordermatrix{ & \\ 00 &+ \f{1}{2} \\ 01 & - \f{1}{2} \\
10 &+ \f{1}{2} \\ 11 & - \f{1}{2} }.$$

We saw from our last attempt at solving this problem, that when we put the bottom qubit into a superposition and
then multiply by $U_f$ we will be in a superposition
$$ (-1)^{f(x)} |x\ra \left[ \f{|0 \ra - |1 \ra}{\sqrt{2}} \right].$$ Now with $|x\ra$ in a superposition,
 we have
$$|\varphi_2 \rangle = \left[ \f{(-1)^{f(0)}|0\rangle + (-1)^{f(1)}|1\rangle}{\sqrt{2}} \right]
\left[\f{|0\rangle - |1\rangle}{\sqrt{2}}\right].$$
Let us carefully look at
$$(-1)^{f(0)}|0\rangle + (-1)^{f(1)|1\rangle }.$$ If $f$ is constant this
becomes either
$$+1(|0\rangle + |1\rangle) \mbox{ or }-1(|0\rangle + |1\rangle)$$
(depending on being constantly $0$ or constantly $1$.) If $f$ is
balanced it becomes either
$$+1(|0\rangle - |1\rangle) \mbox{ or }-1(|0\rangle - |1\rangle)$$
(depending on which way it is balanced.) Summing up, we have that
$$|\varphi_2 \rangle =
 \left\{ \begin{array}{r @{\qquad}l}
(\pm 1) \left[\f{|0\rangle +
|1\rangle}{\sqrt{2}}\right]\left[\f{|0\rangle -
|1\rangle}{\sqrt{2}}\right]
& \mbox{ if $f$ is constant }  \\ \\
(\pm 1)\left[\f{|0\rangle -
|1\rangle}{\sqrt{2}}\right]\left[\f{|0\rangle -
|1\rangle}{\sqrt{2}}\right] & \mbox{ if $f$ is balanced }
\end{array}\right.
$$

Remembering that the Hadamard matrix is its own inverse and takes
$\f{|0\rangle +|1\rangle}{\sqrt{2}}$ to $|0\ra$ and takes
$\f{|0\rangle -|1\rangle}{\sqrt{2}}$ to $|1\ra$, we apply the
Hadamard matrix to the top qubit to get
$$|\varphi_3 \rangle =
 \left\{ \begin{array}{r @{\qquad}l}
(\pm 1)|0\rangle \left[\f{|0\rangle - |1\rangle}{\sqrt{2}}\right] & \mbox{ if $f$ is constant }  \\ \\
(\pm 1)|1\rangle \left[\f{|0\rangle - |1\rangle}{\sqrt{2}}\right] &
\mbox{ if $f$ is balanced } \end{array}\right.
$$
Now we simply measure the top qubit. If it is in state $|0\ra$, then
we know that $f$ is a constant function, otherwise it is a balanced
function. And we did this all with only one evaluation as opposed to
the two evaluations that a classical algorithm demands.

Notice that although the $\pm 1$ tells us even more information,
namely which of the two balanced functions or which of the two
constant functions, we are not privy to this information. Upon
measuring, if the function is balanced, we will measure $|1\ra$
regardless if the state was $(-1)|1\ra$ or $(+1)|1\ra$.

In conclusion, we have solved a problem that a classical computer would require
two function evaluations. Deutsch's quantum algorithm solved the same problem with one
function evaluation. Other quantum algorithms are built-up with similar ideas to the
ones presented here.

\bibliographystyle{plain}
\bibliography{qcbib}

\vspace{.2in}

\noindent Department of Computer and Information Science,\\
Brooklyn College, CUNY,\\
Brooklyn, N.Y. 11210.\\
\\
and\\
\\
Computer Science Department,\\
The Graduate Center, CUNY,\\
New York, N.Y. 10016.\\
\\
e-mail: noson@sci.brooklyn.cuny.edu
\end{document}